\documentclass[conference]{IEEEtran}
\IEEEoverridecommandlockouts
\usepackage{cite}
\usepackage{amsmath,amssymb,amsfonts}
\usepackage{graphicx}
\usepackage{textcomp}
\usepackage{xcolor}
\usepackage{url}
\usepackage{booktabs}
\usepackage{siunitx}
\usepackage{caption}
\usepackage{subcaption}

\def\BibTeX{{\rm B\kern-.05em{\sc i\kern-.025em b}\kern-.08em
    T\kern-.1667em\lower.7ex\hbox{E}\kern-.125emX}}

\usepackage[colorlinks=true,
            linkcolor=black,
            citecolor=black,
            urlcolor=black]{hyperref}

\newcommand{\arxivnotice}{%
Accepted for presentation at the 23rd International Bhurban Conference on
Applied Sciences and Technology (IBCAST 2026), Islamabad, Pakistan,
17--20 August 2026. This is the authors' accepted version; the version of
record will appear in the conference proceedings.
\copyright~2026 IEEE. Personal use of this material is permitted. Permission
from IEEE must be obtained for all other uses, in any current or future media,
including reprinting/republishing this material for advertising or promotional
purposes, creating new collective works, for resale or redistribution to servers
or lists, or reuse of any copyrighted component of this work in other works.
}

\begin{document}

\title{An X-Band Monopulse Direction-Finding Receiver\\
Based on a Rat-Race Comparator and a\\
$2\times2$ Antipodal Vivaldi Array}

\author{
  \IEEEauthorblockN{M. Ahmad Nabil, Sohaib Yaqoob Chaudhry, Salman Liaquat}
  \IEEEauthorblockA{National University of Science and Technology, Pakistan\\syaqoob@cae.nust.edu.pk}
  \thanks{\arxivnotice}
}

\maketitle

\begin{abstract}
This paper presents the design, fabrication, and experimental validation of a compact monopulse direction-finding (DF) receiver operating in the X-band. The receiver combines a $2 \times 2$ antipodal Vivaldi antenna array with a rat-race (\SI{180}{\degree}-hybrid) comparator network that simultaneously synthesizes the sum ($\Sigma$), azimuth-difference ($\Delta_{\text{az}}$), and elevation-difference ($\Delta_{\text{el}}$) channels required for monopulse processing, removing the need for mechanical scanning. The Vivaldi element provides good impedance matching across \SIrange{8}{12}{\giga\hertz} with a simulated realized gain of approximately \SI{6.44}{dBi}, and the comparator exhibits the \SI{180}{\degree} phase balance at its difference ports required for angle encoding, with the fabricated prototype preserving the matched, balanced response measured across the band. Three LTC5564 envelope detectors convert the channel outputs to DC voltages that are digitized by an on-board microcontroller and processed in software to form the monopulse ratios and estimate the direction of arrival (DoA), with the result displayed in real time on a MATLAB graphical user interface. Azimuth direction finding is experimentally demonstrated over fourteen angles spanning \SIrange{-20}{19}{\degree} at \SI{11}{\giga\hertz}, yielding a root-mean-square error (RMSE) of \SI{7}{\degree}; the architecture is directly extensible to elevation. The result is a passive, low-cost, and fully integrated receiver suitable for radar sensing, electronic warfare, and surveillance.
\end{abstract}


\section{Introduction}

The ability of direction-finding (DF) systems to accurately determine the angular location of a target makes them indispensable in modern applications such as surveillance, aviation, wireless communications, electronic warfare, autonomous platforms, and military operations \cite{ref1,ref2}. Airborne early-warning systems, missile guidance, target tracking, and phased-array radars all demand accurate, near-real-time angle estimation with low latency and high angular resolution \cite{ref3}.
 
Among the techniques used for angle estimation, monopulse radar is one of the most effective because it provides instantaneous angular information from a single received pulse \cite{ref4}. Compared with conventional sequential-scanning methods, monopulse offers improved tracking accuracy, faster response, and reduced angular ambiguity \cite{ref5}. A monopulse system forms a sum ($\Sigma$) signal and one or more difference ($\Delta$) signals; the ratio of the difference signal to the sum signal is monotonic in arrival angle and is therefore used to estimate angular displacement \cite{ref4}.
 
Conventional monopulse radars typically employ reflector or horn antennas with mechanically scanned configurations, which result in bulky structures, significant weight, and complex feeding networks \cite{ref6}. Planar antennas have consequently gained attention owing to their compact size, low fabrication cost, and compatibility with microwave integrated circuits \cite{ref7,ref8}, motivating extensive work on planar monopulse antennas, comparator architectures, and DF techniques \cite{ref10,ref11}.
 
The comparator network is among the most critical components of a monopulse DF system, since it must synthesise the sum and difference channels while preserving the required amplitude and phase balance \cite{ref12}. Several comparator topologies exist; branch-line and rat-race (\SI{180}{\degree}-hybrid) couplers are the most widely used \cite{ref8}. Rat-race comparators are attractive because of their symmetric architecture, reciprocal operation, good port isolation, and ease of integration in planar technology \cite{ref14}. Recent work has continued to refine planar rat-race comparator networks: a wideband design realised in inverted-microstrip gap-waveguide technology achieves close to \SI{50}{\percent} bandwidth from four rat-race hybrids \cite{ref20}, while a fully symmetric uni-planar comparator network has been proposed to overcome the port-crossing and amplitude-imbalance limitations of the conventional \SI{180}{\degree} coupler \cite{ref21}. These efforts confirm that the rat-race remains the workhorse of planar monopulse feeding networks.

Among broadband radiators, antipodal Vivaldi antennas are well suited to radar and DF applications owing to their ultra-wide bandwidth, high gain, stable radiation characteristics, and suitability for array integration \cite{ref3,ref7}. They have demonstrated strong potential in phased arrays, microwave imaging, beamforming, and wideband sensing \cite{ref16}. Recent designs have further improved their gain and directivity for radar and tracking use \cite{ref22}, and tightly-coupled antipodal Vivaldi arrays have been shown to extend the lower band edge while retaining the wide bandwidth required for accurate DoA estimation \cite{ref24}.
 
Despite this progress, most reported monopulse studies validate either the comparator network or the antenna in isolation, and many remain simulation-only. Comparatively few works demonstrate a fully assembled, low-cost receiver in which a planar antenna array, a rat-race comparator, an envelope-detector front-end, and a software DoA estimator are integrated and measured end-to-end. This paper addresses that gap by presenting a complete, experimentally validated monopulse DF receiver that combines a $2 \times 2$ antipodal Vivaldi array with a rat-race comparator network operating across the X-band. The comparator generates three channels---$\Sigma$, $\Delta_{\text{az}}$, and $\Delta_{\text{el}}$---which are envelope-detected by LTC5564 power detectors, digitised, and processed in software to estimate the DoA, with results displayed in real time through a MATLAB graphical user interface (GUI). The remainder of the paper is organised as follows. Section~\ref{sec:method} describes the proposed methodology, covering the system architecture, antenna and comparator design, power-detector design, and the DoA estimation algorithm. Section~\ref{sec:results} presents the simulated and measured results together with the experimental DF performance. Section~\ref{sec:conclusion} concludes the paper.

\begin{figure}[t]
  \centering
  \includegraphics[width=\columnwidth]{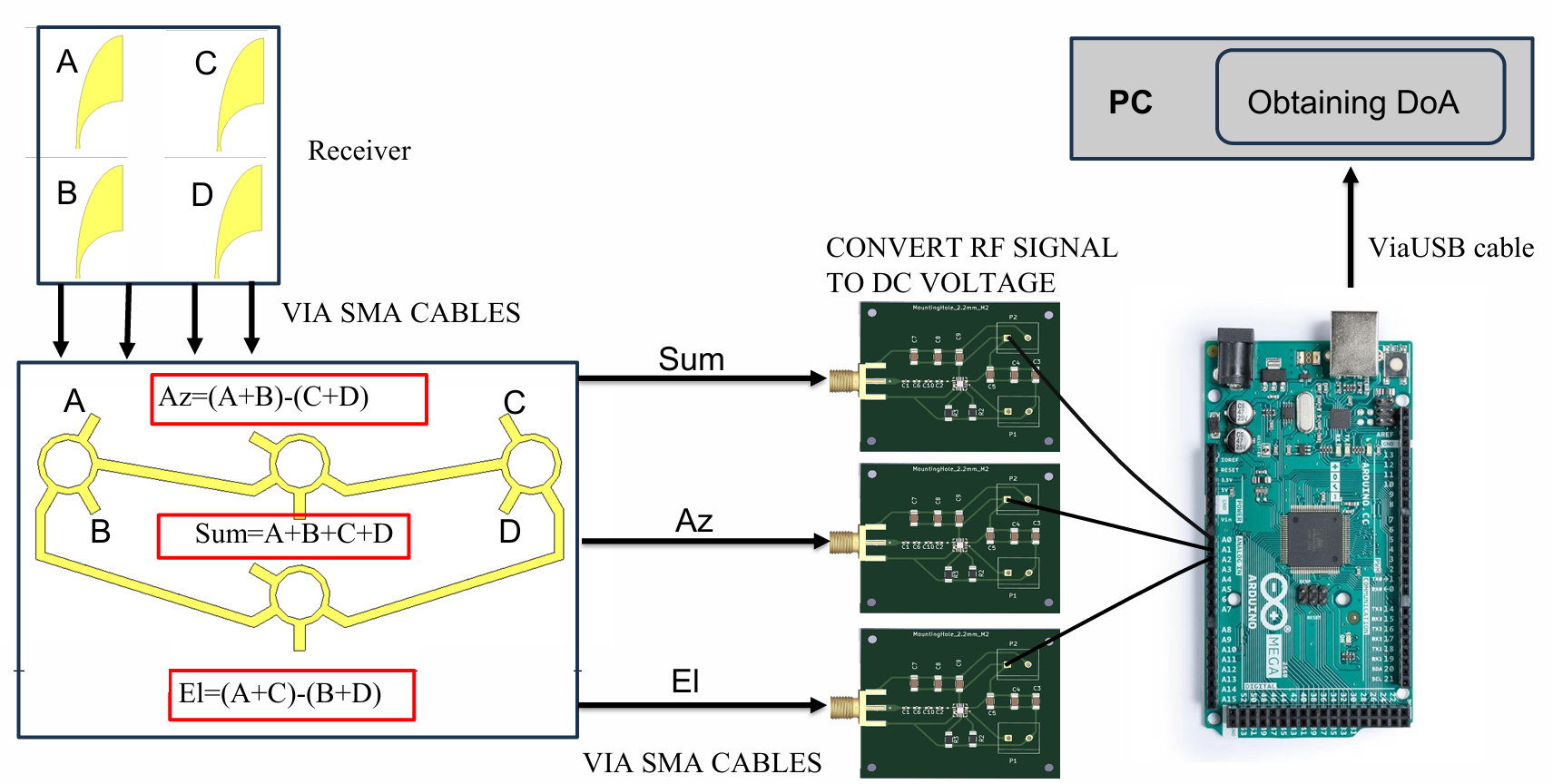}
  \caption{System block diagram of the proposed monopulse direction-finding receiver.}
  \label{fig:block}
\end{figure}

\begin{figure}[t]
  \centering
  \begin{subfigure}{0.48\columnwidth}
    \centering
    \includegraphics[width=\textwidth,height=4cm,keepaspectratio]{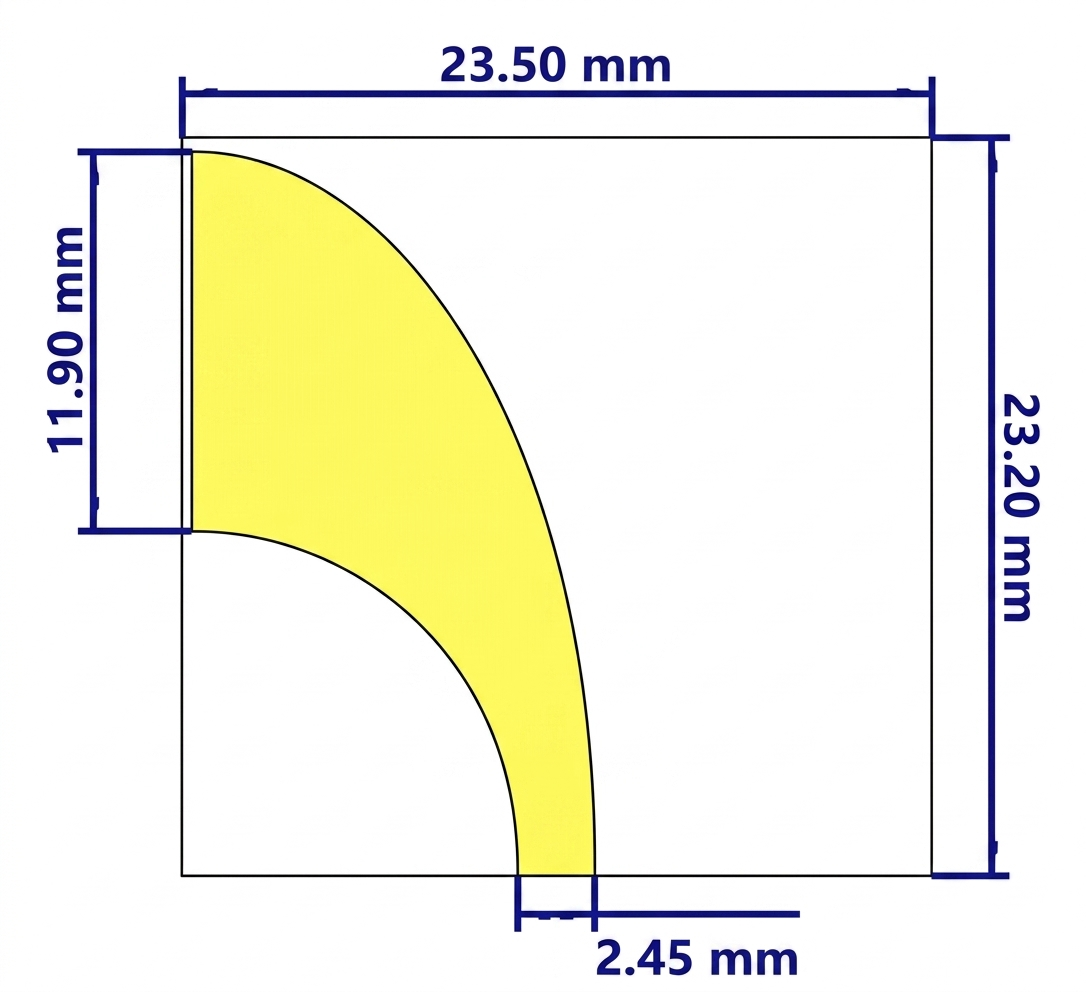}
    \caption{}
    \label{fig:vivaldi_elem}
  \end{subfigure}
  \hfill
  \begin{subfigure}{0.48\columnwidth}
    \centering
    \includegraphics[width=\textwidth,height=4cm,keepaspectratio]{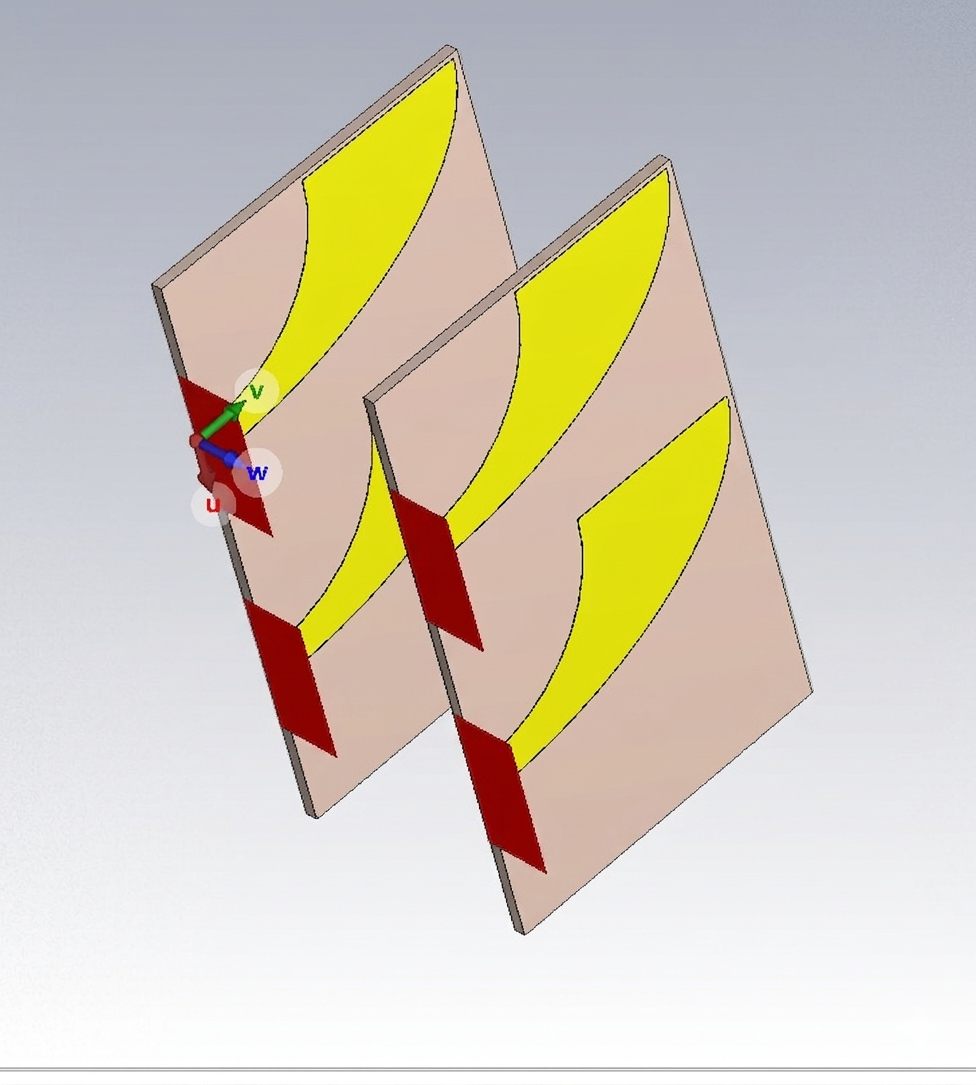}
    \caption{}
    \label{fig:vivaldi_arr}
  \end{subfigure}
  \caption{Antipodal Vivaldi antenna: (a) single element and (b) $2 \times 2$ array configuration.}
  \label{fig:vivaldi}
\end{figure}
\section{Proposed Methodology}
\label{sec:method}

\subsection{System Architecture and Operating Principle}
The system architecture is shown in Fig.~\ref{fig:block}. A $2 \times 2$ planar array of antipodal Vivaldi elements ($A$--$D$) forms the receiving aperture, capturing the incident wavefront in both the azimuth and elevation planes. The four element outputs feed a rat-race comparator network (\SI{180}{\degree}-hybrid), which performs analog beamforming to synthesize three monopulse channels: a sum channel $\Sigma = A + B + C + D$, which serves as the amplitude reference; an azimuth-difference channel $\Delta_{\text{az}} = (A + B) - (C + D)$; and an elevation-difference channel $\Delta_{\text{el}} = (A + C) - (B + D)$. The two difference channels encode the angular displacement of the source in orthogonal planes.

Each channel is envelope-detected by an LTC5564 power detector, producing a DC voltage proportional to the channel power. These voltages are digitized by an on-board analog-to-digital converter (ADC) and streamed to a host PC, where the monopulse ratios $\Delta_{\text{az}}/\Sigma$ and $\Delta_{\text{el}}/\Sigma$ are formed and mapped to azimuth and elevation estimates. Normalizing the difference channels by $\Sigma$ removes the dependence on absolute received power, leaving a quantity that varies monotonically with the arrival angle. 

\subsection{Antenna Array Design}
\begin{figure}[t]
  \centering
  \includegraphics[width=0.9\columnwidth,height=0.22\textheight,keepaspectratio]{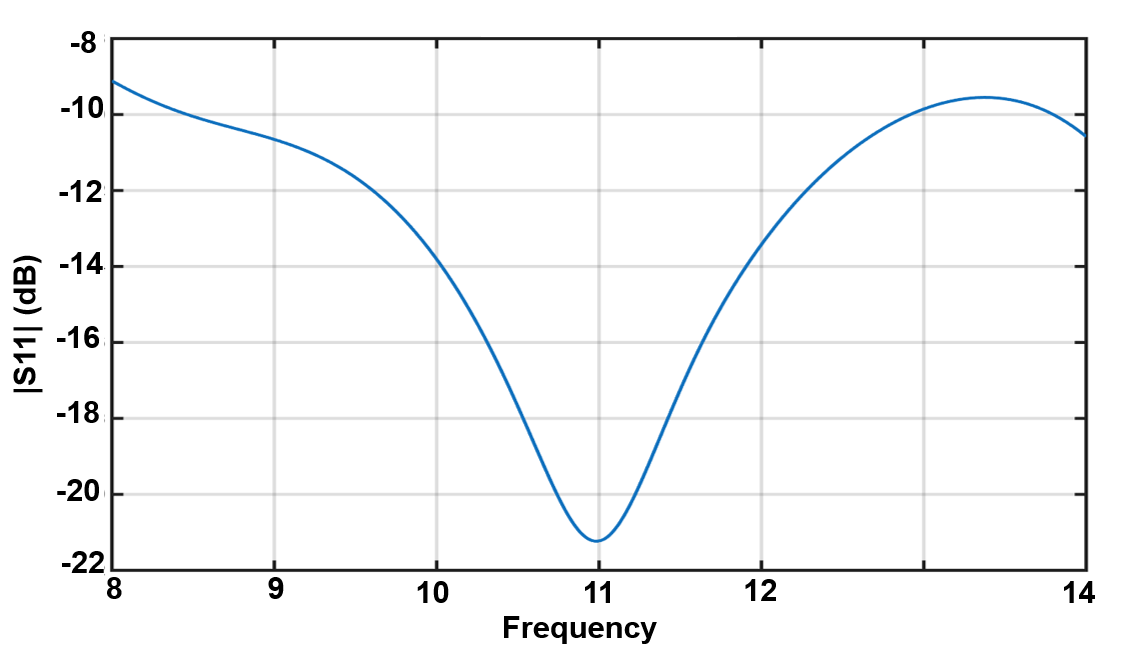}
  \caption{Simulated $S_{11}$ of the antipodal Vivaldi antenna element.}
  \label{fig:s11}
\end{figure}

The antenna was designed and optimized in CST Microwave Studio. An antipodal single Vivaldi element operating in the X-band was initially designed (Fig.~\ref{fig:vivaldi}). The substrate is Rogers RO4003C ($\varepsilon_r = 3.55$), chosen for its low loss at X-band frequencies. The simulated $S_{11}$ confirms good matching across the X-band, with an impedance bandwidth spanning \SIrange{8}{12}{\giga\hertz} (Fig.~\ref{fig:s11}). The four identical elements were then arranged into a $2 \times 2$ planar array (Fig.~\ref{fig:vivaldi}), which forms the receiving front-end of the DoA estimation system.

\begin{figure}[t]
  \centering
  \includegraphics[width=0.8\columnwidth]{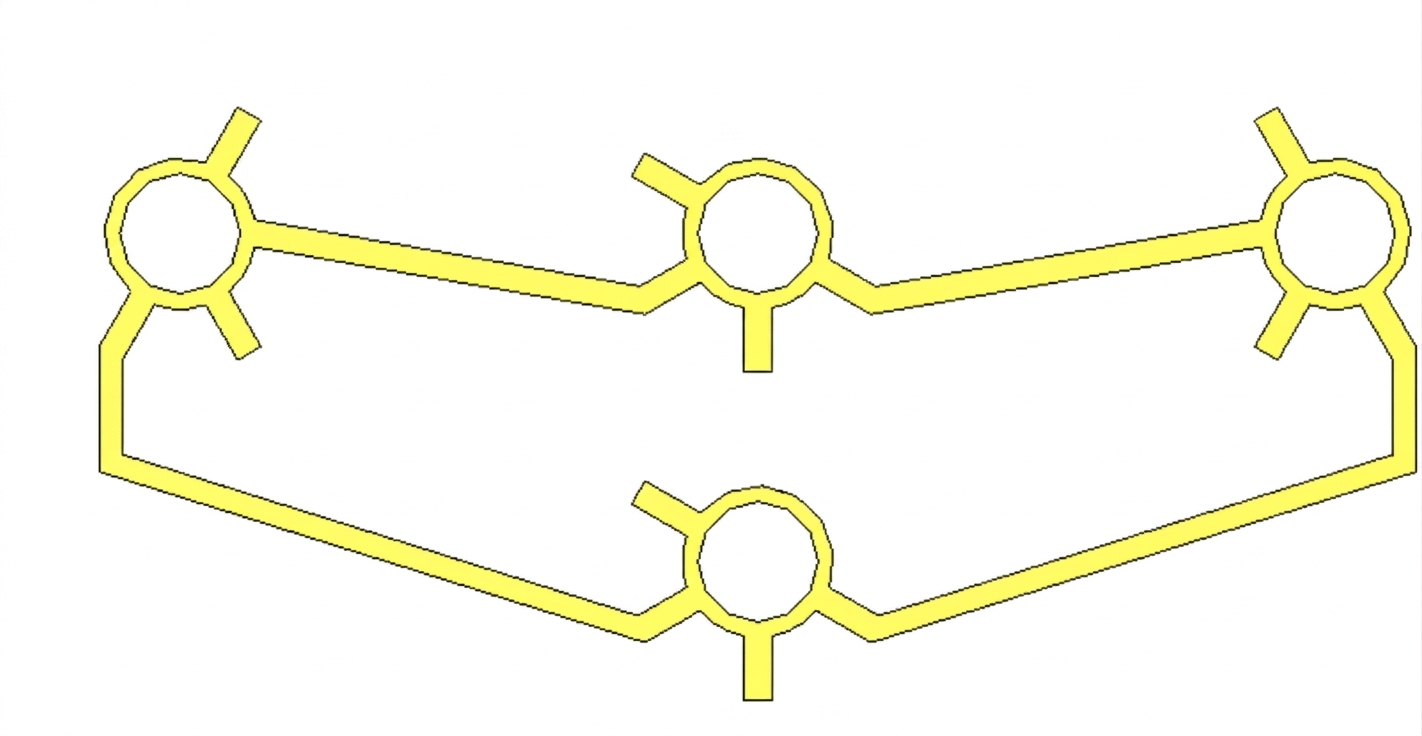}
  \caption{Rat-race comparator network formed by four interconnected \SI{180}{\degree}-hybrid couplers.}
  \label{fig:ratrace}
\end{figure}

\begin{figure}[t]
  \centering
  \includegraphics[width=0.75\columnwidth,height=0.22\textheight,keepaspectratio]{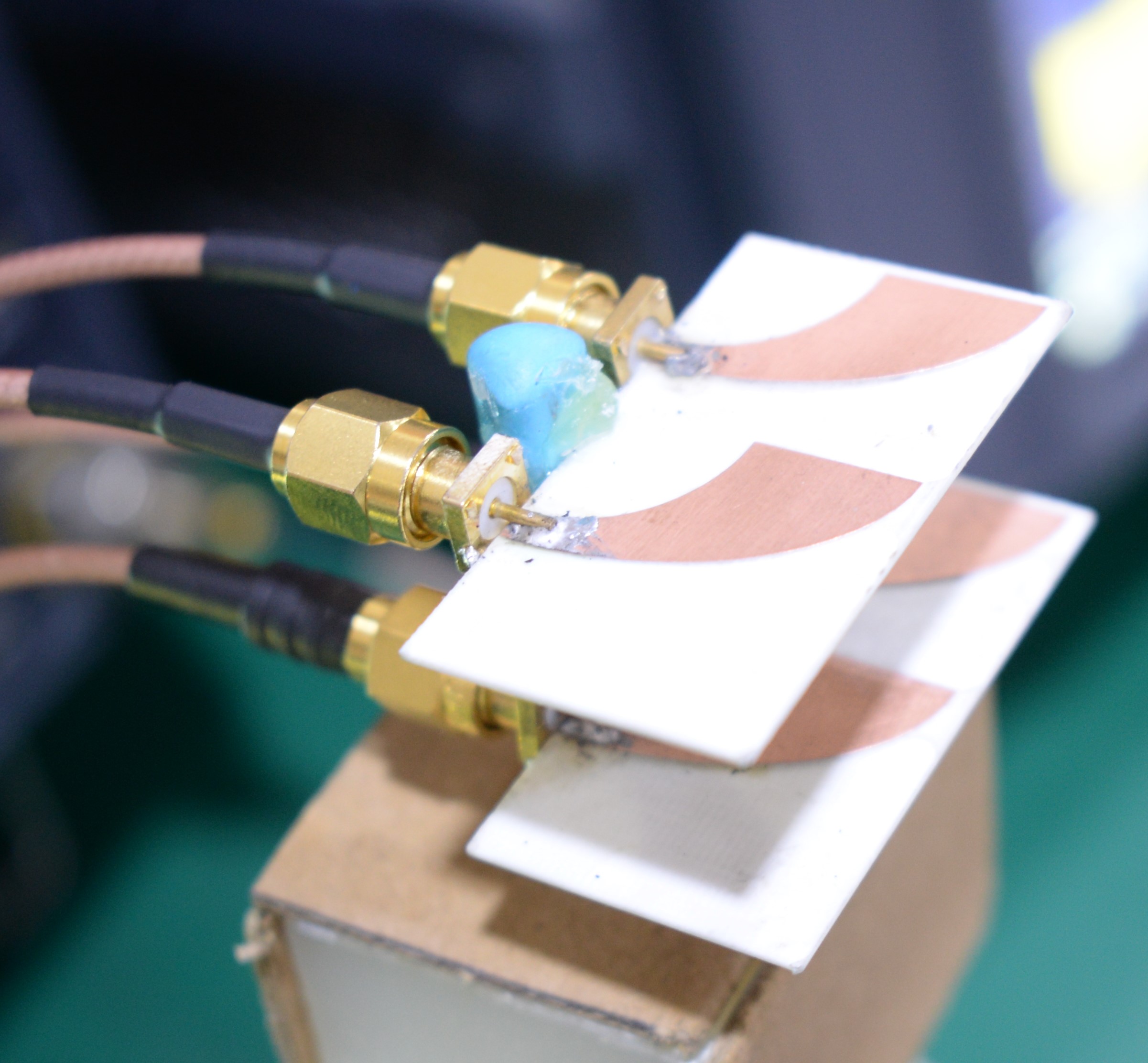}
  \caption{Fabricated $2 \times 2$ antipodal Vivaldi antenna array.}
  \label{fig:fab_array}
\end{figure}

\begin{figure}[t]
  \centering
  \includegraphics[width=0.85\columnwidth,height=0.22\textheight,keepaspectratio]{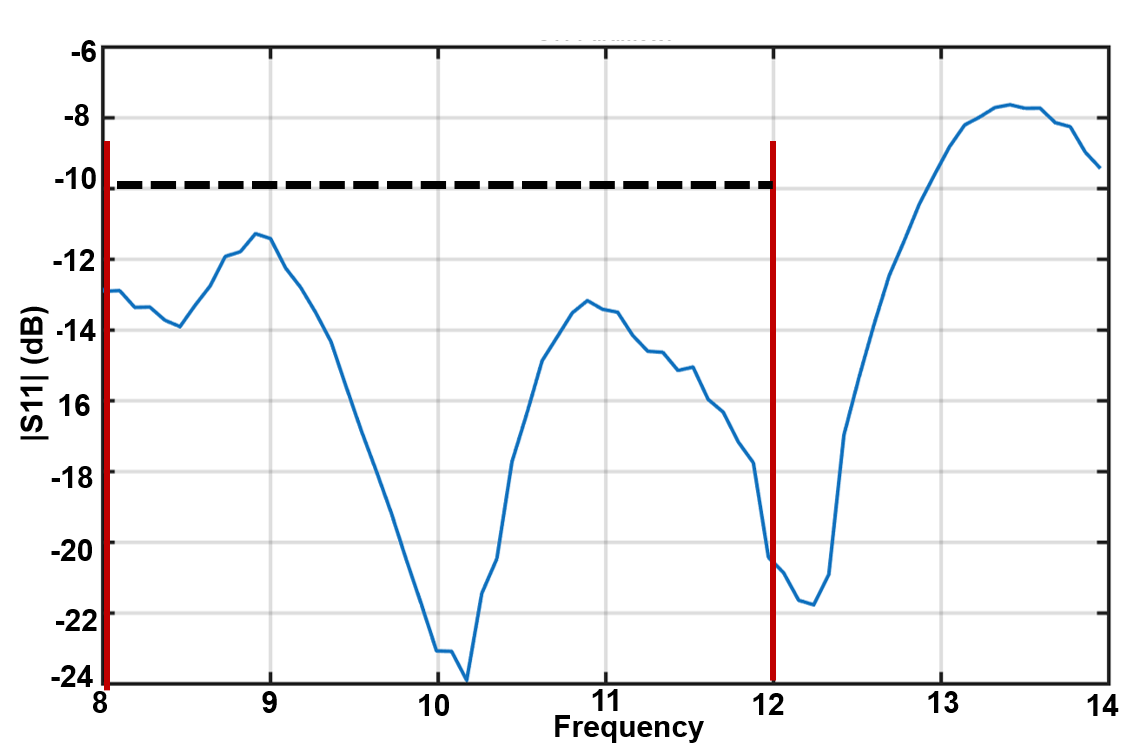}
  \caption{Measured $S_{11}$ of the fabricated antenna array.}
  \label{fig:meas_array}
\end{figure}

\begin{figure}[t]
  \centering
  \includegraphics[width=\columnwidth]{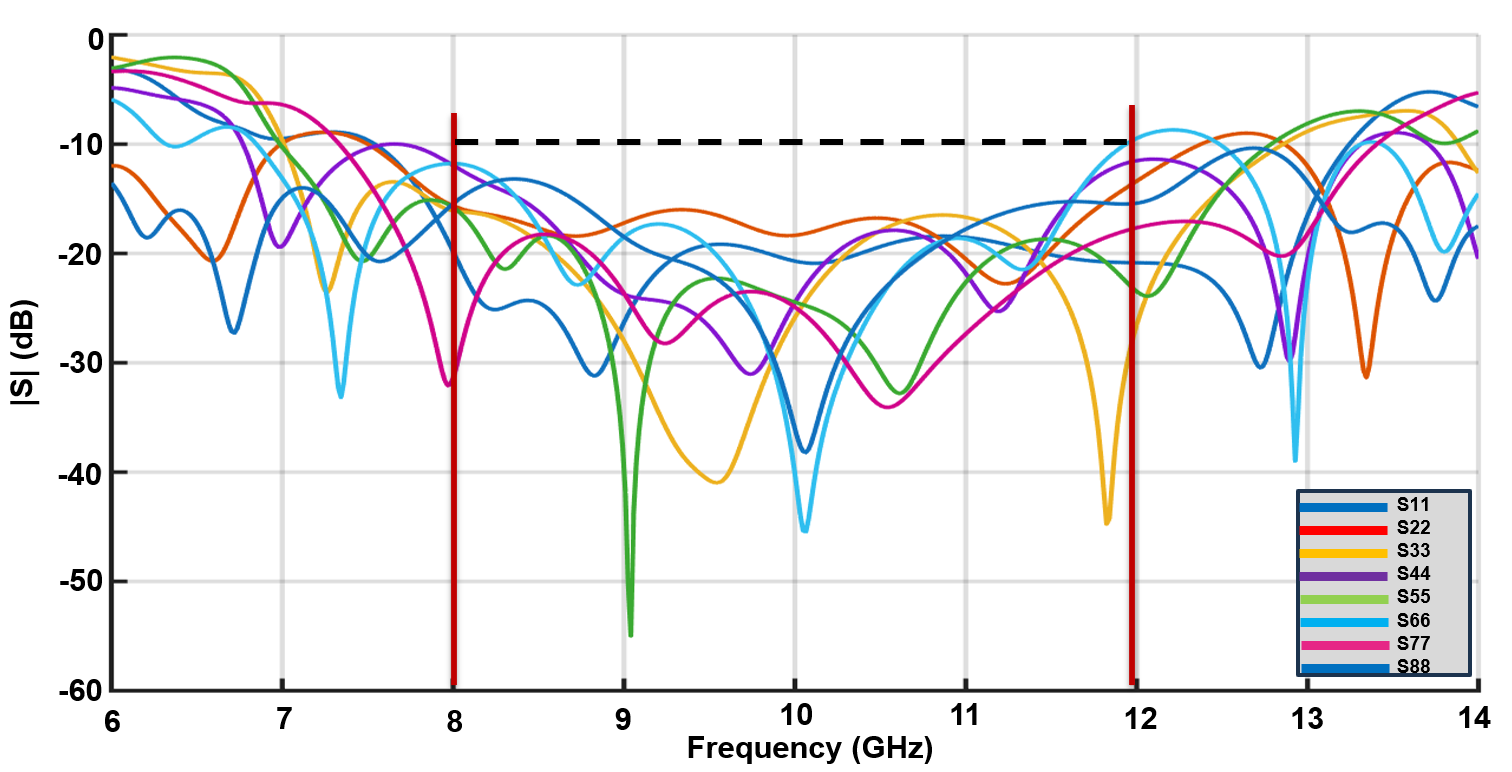}
  \caption{Simulated scattering parameters of the comparator network.}
  \label{fig:comp_sim}
\end{figure}

\begin{figure}[t]
  \centering
  \includegraphics[width=\columnwidth]{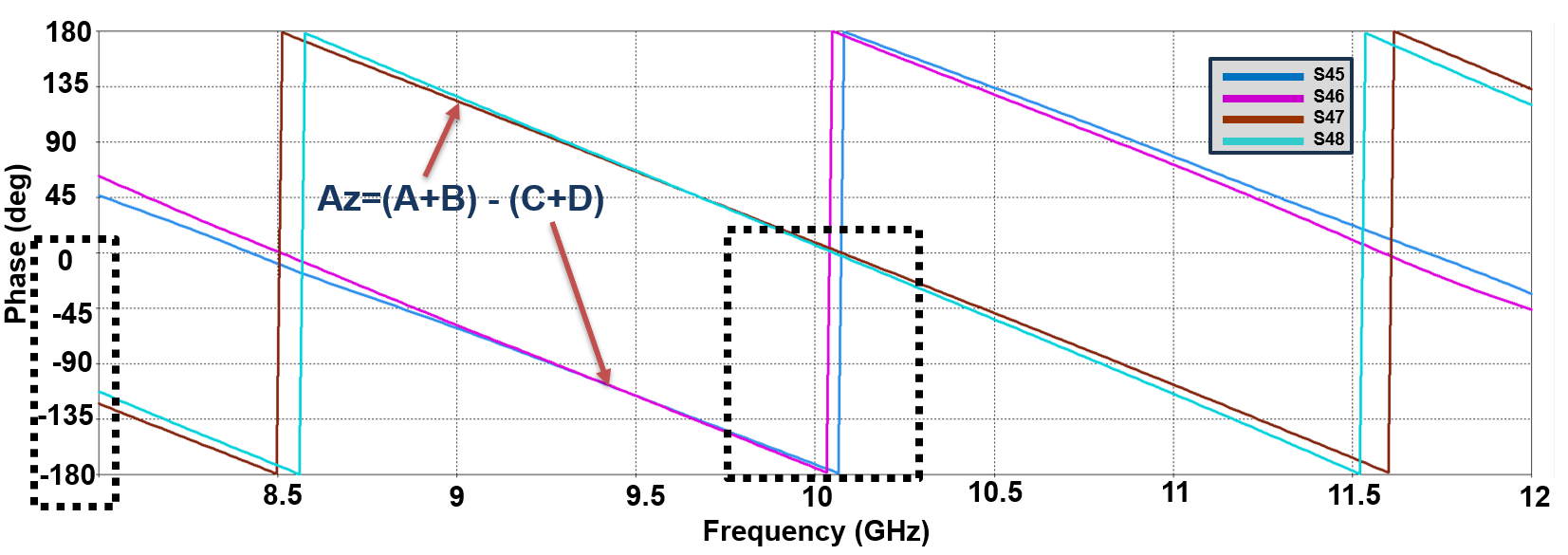}
  \caption{Simulated \SI{180}{\degree} phase difference at the azimuth-difference ($\Delta_{\text{az}}$) port.}
  \label{fig:phase_az}
\end{figure}
\subsection{Rat-Race Comparator Design}
For a two-dimensional DF, four rat-race (\SI{180}{\degree}-hybrid) couplers are interconnected to form the comparator network shown in Fig.~\ref{fig:ratrace}, which produces the three-channel sum--difference output from the $2 \times 2$ array. Each ring coupler was designed with an outer radius of \SI{4.85}{\milli\meter} and an inner radius of \SI{3.85}{\milli\meter}, with all ports referenced to a characteristic impedance of \SI{50}{\ohm}. The network splits and recombines the element signals to produce the $\Sigma$, $\Delta_{\text{az}}$, and $\Delta_{\text{el}}$ channels; forming $\Delta_{\text{az}}/\Sigma$ and $\Delta_{\text{el}}/\Sigma$ that gives the instantaneous azimuth and elevation angles, without any mechanical scanning.

\subsection{X-Band Power-Detector PCB}
To convert RF channel outputs into proportional DC voltages, a custom power-detector printed circuit board (PCB) was designed in KiCad using the LTC5564 RF power-detector IC, which operates in the X-band. Three identical detector modules process the outputs $\Sigma$, $\Delta_{\text{az}}$, and $\Delta_{\text{el}}$, each producing a DC voltage proportional to the incident RF power on its channel.


\subsection{Direction-Finding Algorithm}
The comparator produces the three main outputs: 
\begin{align}
  \Sigma &= A + B + C + D, \label{eq:sigma}\\
  \Delta_{\mathrm{az}} &= (A + B) - (C + D), \label{eq:delta_az}\\
  \Delta_{\mathrm{el}} &= (A + C) - (B + D). \label{eq:delta_el}
\end{align}
The azimuth and elevation monopulse ratios are defined as
\begin{equation}
  R_{\mathrm{az}} = \frac{\Delta_{\mathrm{az}}}{\Sigma}, \qquad
  R_{\mathrm{el}} = \frac{\Delta_{\mathrm{el}}}{\Sigma},
  \label{eq:ratios}
\end{equation}
and the corresponding angles are estimated through
\begin{equation}
  \theta_{\mathrm{az}} \approx K_{\mathrm{az}}\,R_{\mathrm{az}}, \qquad
  \theta_{\mathrm{el}} \approx K_{\mathrm{el}}\,R_{\mathrm{el}},
  \label{eq:theta}
\end{equation}
where $K_{\mathrm{az}}$ and $K_{\mathrm{el}}$ are calibration constants. Because the monopulse ratio is not perfectly linear over the full angular range, the linear relationship in \eqref{eq:theta} is replaced in the implementation by a measured lookup table (LUT) that maps each ratio value to its corresponding angle, improving estimation accuracy near the beam edges.

\section{Results and Discussion}
\label{sec:results}

\subsection{Simulated and Measured Antenna Performance}
The simulated $S_{11}$ of the Vivaldi element (Fig.~\ref{fig:s11}) confirms good impedance matching across \SIrange{8}{12}{\giga\hertz}. The fabricated $2 \times 2$ array is shown in Fig.~\ref{fig:fab_array}, with each element connected to the comparator network. The measured $S_{11}$ (Fig.~\ref{fig:meas_array}) agrees well with simulation, retaining sub-\SI{-10}{dB} matching across the X-band.


\subsection{Comparator Performance}
\begin{figure}[t]
  \centering
  \begin{subfigure}{0.45\columnwidth}
    \centering
    \includegraphics[width=\textwidth,height=4cm,keepaspectratio]{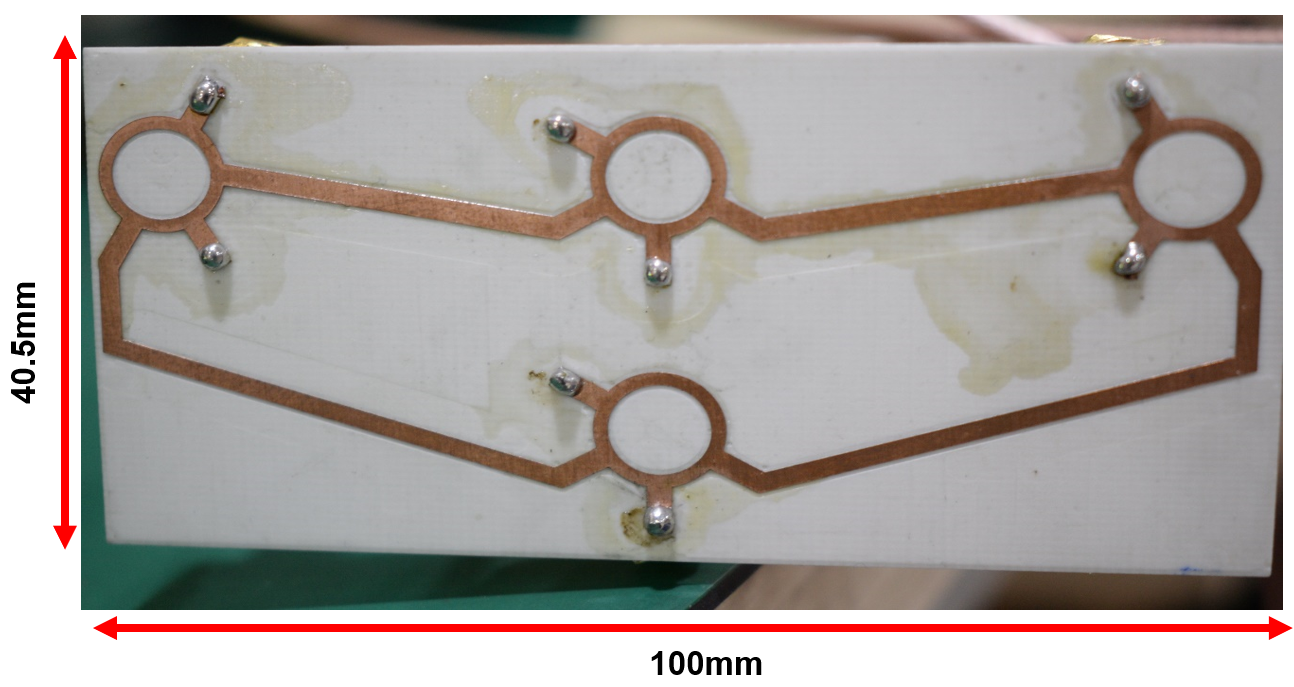}
    \caption{}
    \label{fig:fab_comp}
  \end{subfigure}
  \hfill
  \begin{subfigure}{0.45\columnwidth}
    \centering
    \includegraphics[width=\textwidth,height=4cm,keepaspectratio]{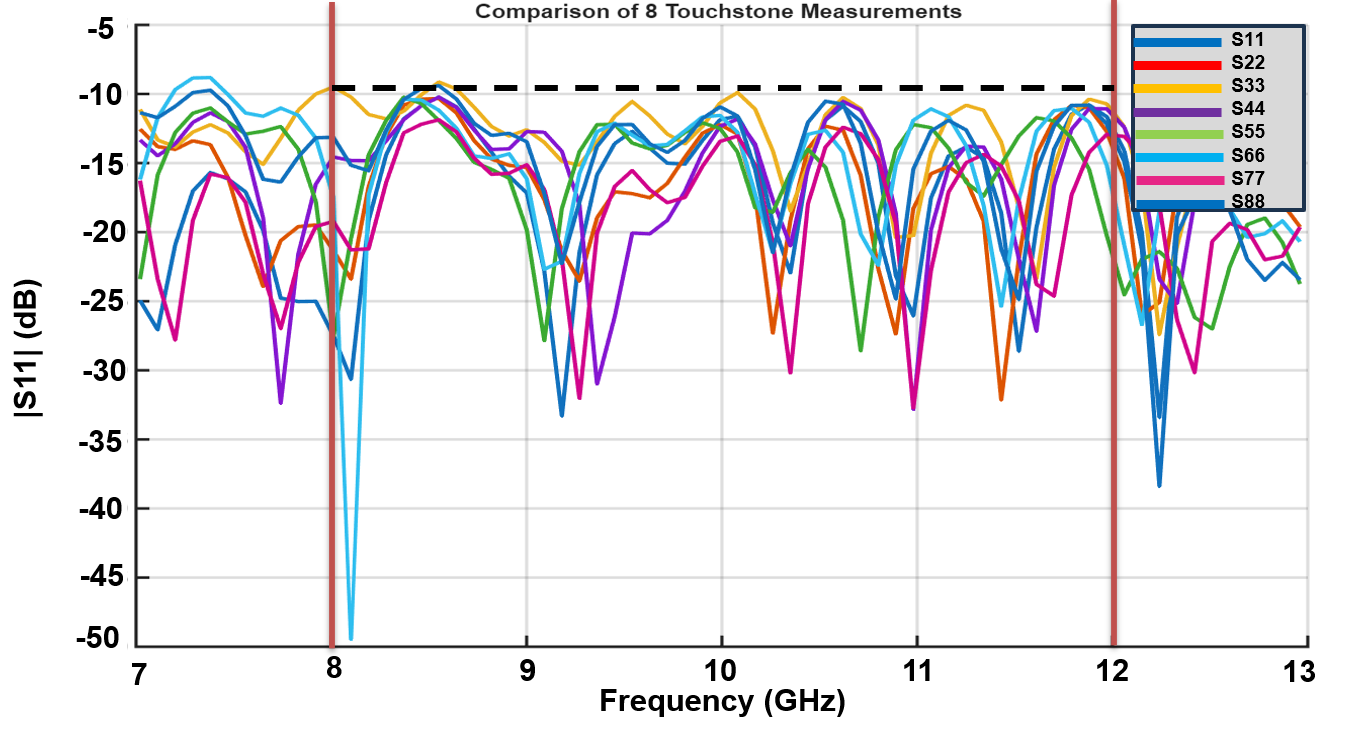}
    \caption{}
    \label{fig:comp_meas}
  \end{subfigure}
  \caption{(a) Rat-race comparator network (b) Measured S-parameters of the comparator}
  \label{fig:vp}
\end{figure}

The simulated scattering parameters of the comparator are shown in Fig.~\ref{fig:comp_sim}, all ports are matched across \SIrange{8}{12}{\giga\hertz} frequency range. The key figure of merit for a monopulse comparator is the phase balance at its output ports. Across the band, both difference ports maintain a consistent \SI{180}{\degree} phase separation between their respective element groups, left/right for $\Delta_{\text{az}}$ (Fig.~\ref{fig:phase_az}) and upper/lower for $\Delta_{\text{el}}$, while the sum port remains in phase. This antiphase relationship at the difference ports is the condition required for correct two-dimensional angle encoding which guarantees a deep null on the boresight axis, which the monopulse ratio exploits to resolve angle. The fabricated comparator (Fig.~\ref{fig:fab_comp}) was characterized through its measured scattering parameters (Fig.~\ref{fig:comp_meas}); the measured response preserves the matched, balanced behavior predicted by simulation across the operating band, with the small deviations attributable to fabrication inaccuracies.



\begin{figure}[t]
  \centering
  \begin{subfigure}{0.48\columnwidth}
    \centering
    \includegraphics[width=\textwidth,height=4cm,keepaspectratio]{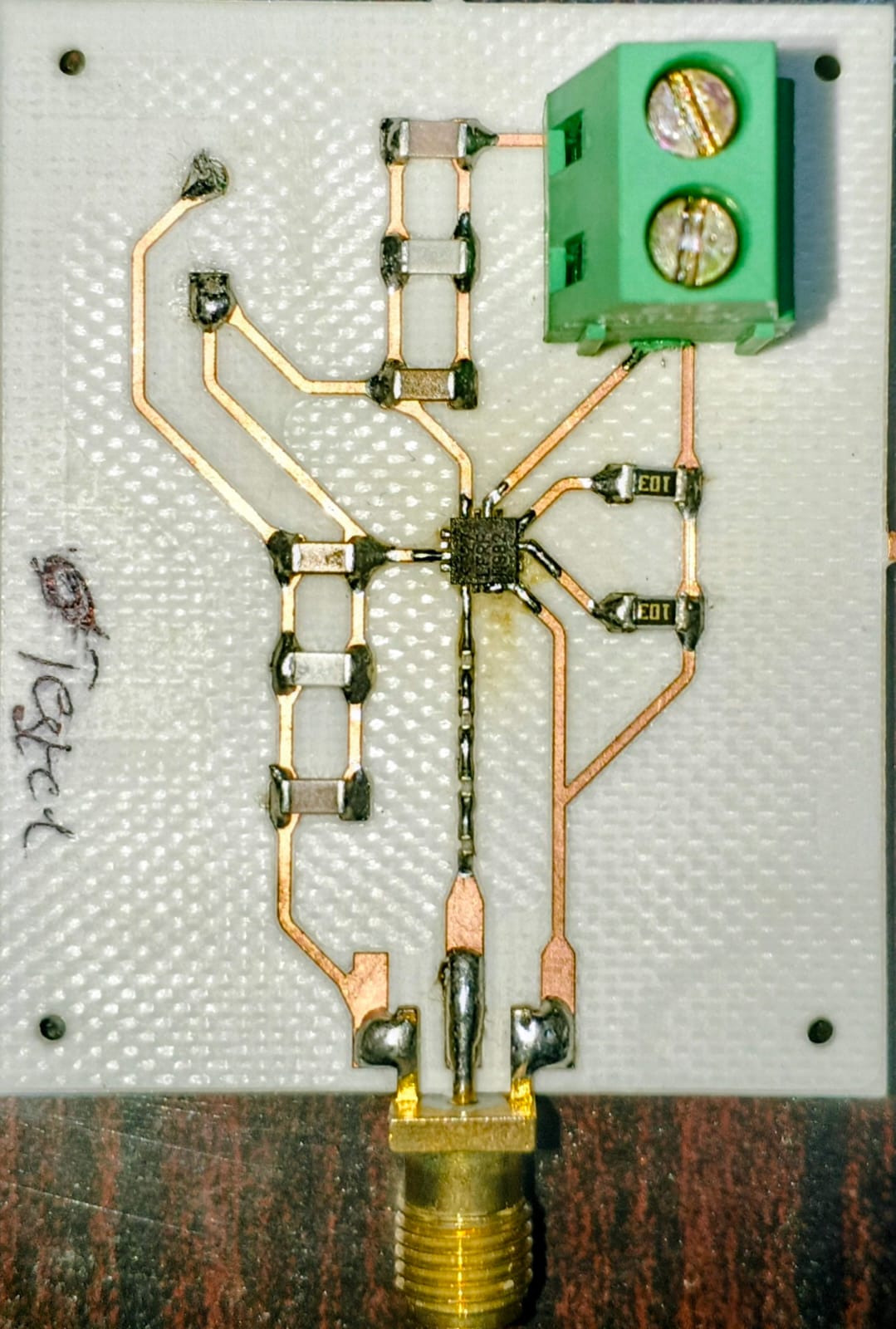}
    \caption{}
    \label{fig:fab_pcb}
  \end{subfigure}
  \hfill
  \begin{subfigure}{0.48\columnwidth}
    \centering
    \includegraphics[width=\textwidth,height=4cm,keepaspectratio]{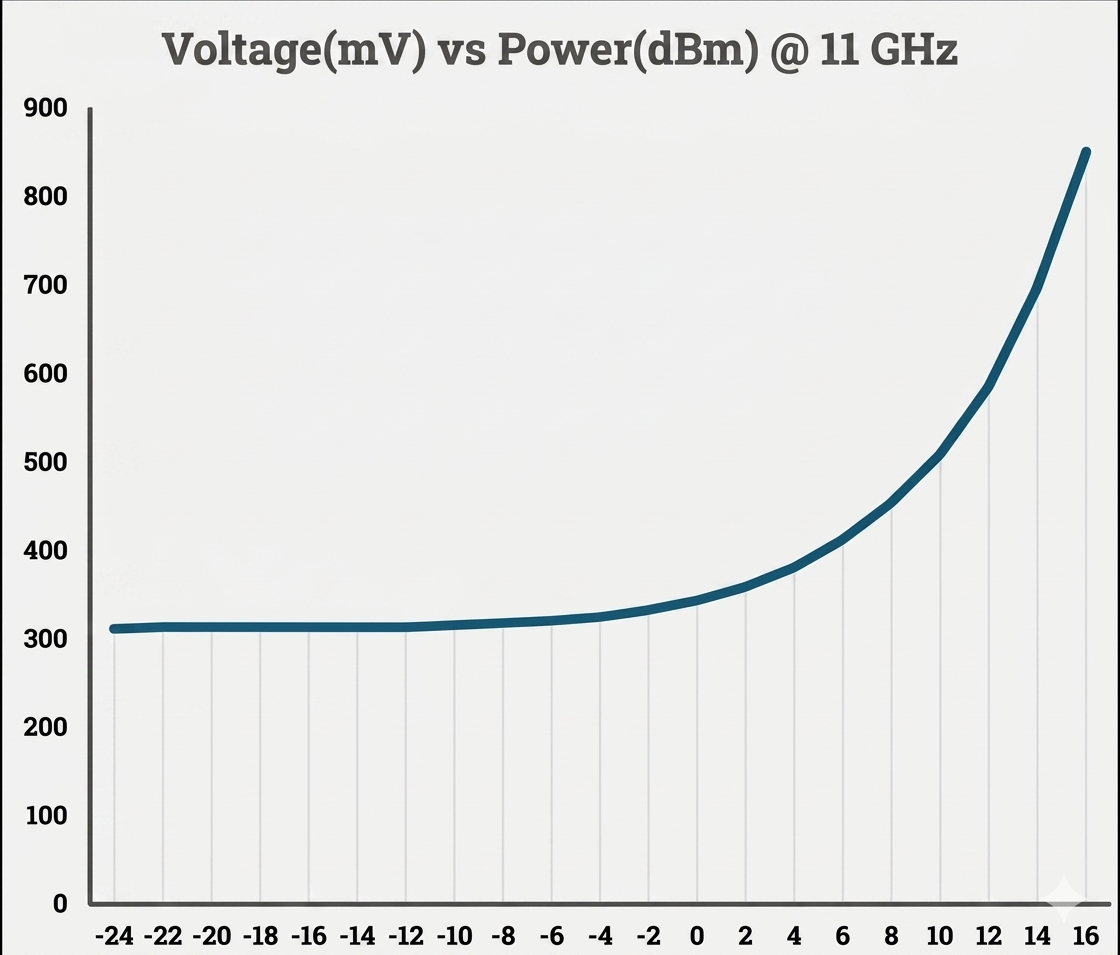}
    \caption{}
    \label{fig:vivaldi_arr}
  \end{subfigure}
  \caption{(a) X-band power-detector PCB (b) Measured output voltage versus input power of the power detector}
  \label{fig:vp}
\end{figure}

\subsection{Power-Detector Response}
The voltage--power characteristic of the X-band power-detector PCB (Fig.~\ref{fig:fab_pcb}) was measured by applying a range of input power levels at \SI{11}{\giga\hertz} and recording the corresponding DC output voltage (Fig.~\ref{fig:vp}). The output voltage increases monotonically with input RF power, confirming the expected detector response and verifying the suitability of the board for monopulse signal processing.

\subsection{Integrated System and Direction-Finding Results}
\begin{figure}[t]
  \centering
  \includegraphics[width=\columnwidth]{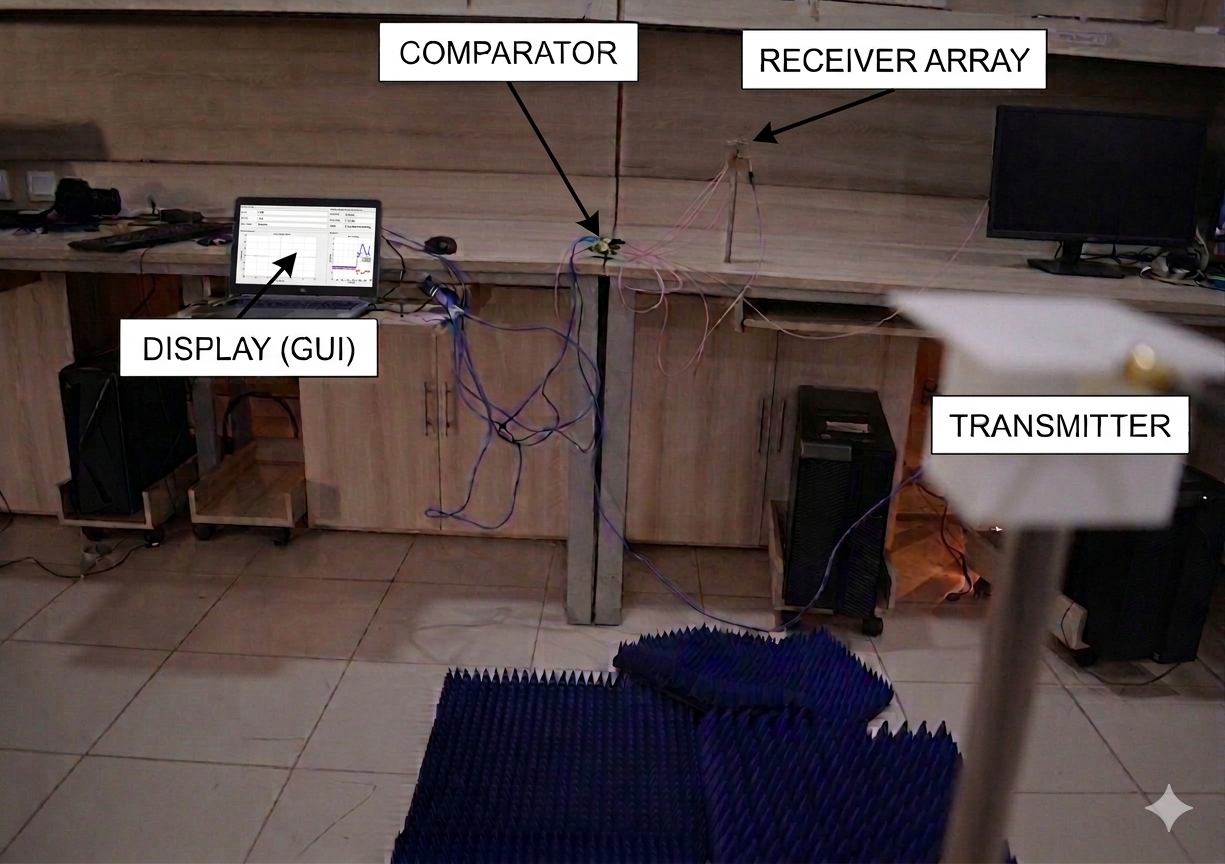}
  \caption{Experimental measurement setup for direction finding.}
  \label{fig:setup}
\end{figure}

\begin{figure}[t]
  \centering
  \includegraphics[width=\columnwidth]{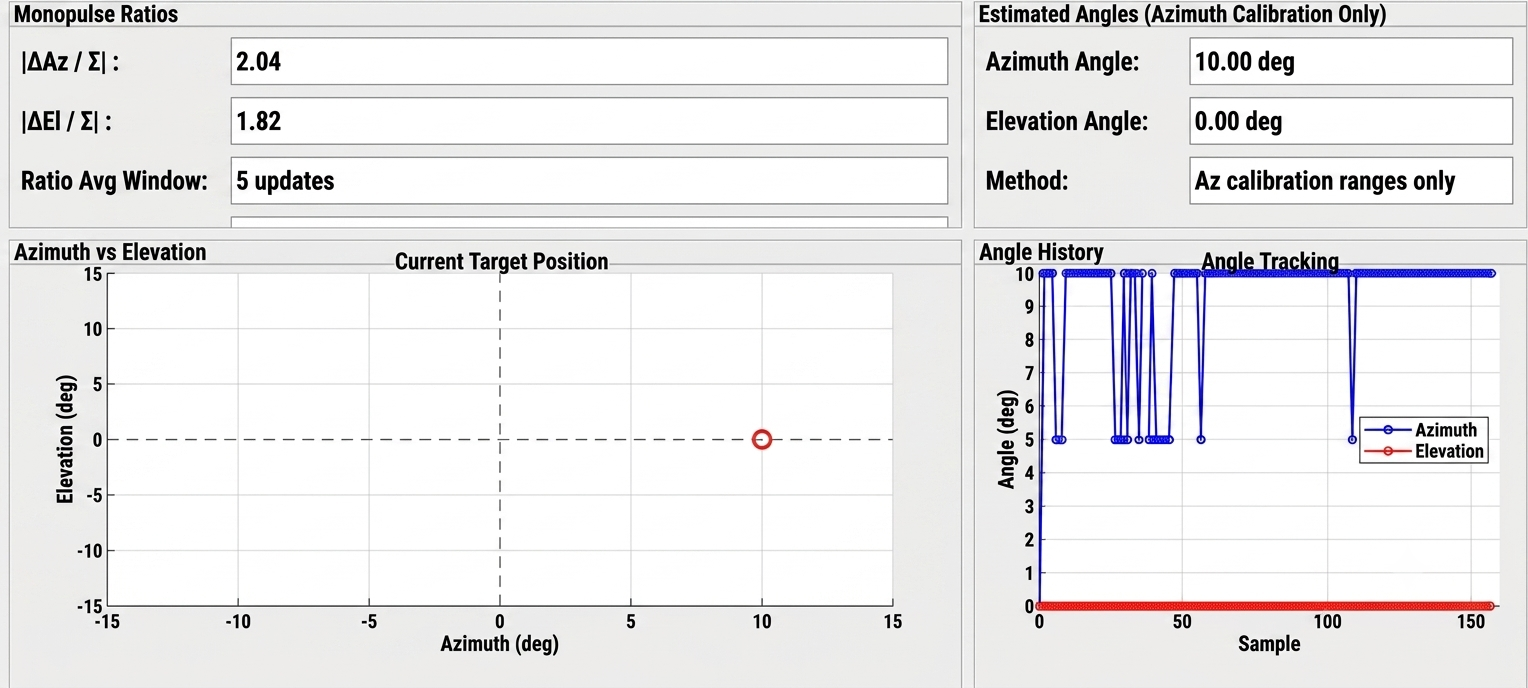}
  \caption{MATLAB graphical user interface displaying the estimated angle plot in real time.}
  \label{fig:gui}
\end{figure}

The integrated measurement setup is shown in Fig.~\ref{fig:setup}. A transmitting antenna placed \SI{2}{\meter} from the receiving array emits an X-band signal; as its angular position is varied, the array captures the incident field and feeds it to the comparator. The three power-detector boards convert the $\Sigma$, $\Delta_{\text{az}}$, and $\Delta_{\text{el}}$ outputs to DC voltages, which are digitized by an Arduino and transmitted over a serial link to a host PC for monopulse-ratio computation and DoA estimation. A measured lookup table relates the azimuth monopulse ratio to angle, and the MATLAB GUI (Fig.~\ref{fig:gui}) displays the estimated angle in real time.

To quantify accuracy, the source was placed at fourteen azimuth angles spanning \SIrange{-20}{19}{\degree}, and the estimated angles were compared against the true positions, as summarised in Table~\ref{tab:az_angles}. The estimation accuracy is given by the root-mean-square error
\begin{equation}
  \mathrm{RMSE} = \sqrt{\frac{1}{N}\sum_{i=1}^{N} e_i^{2}},
  \label{eq:rmse}
\end{equation}
where $e_i$ is the per-sample estimation error and $N$ is the number of test angles. Over the $N = 14$ measured angles, the system achieved an azimuth-estimation RMSE of \SI{7}{\degree}. As Table~\ref{tab:az_angles} shows, the per-angle error remains bounded within $\pm\SI{9}{\degree}$ and is smallest near boresight, where the monopulse ratio is steepest and most sensitive to angle; the larger deviations occur toward the edges of the field of view, consistent with the flattening of the ratio curve at wide angles and with the residual multipath in the laboratory environment. The overall accuracy confirms reliable single-pulse angle estimation across the tested range.

\begin{table}[h]
\centering
\caption{True vs. Estimated Azimuth Angles (RMSE = $7^\circ$)}
\label{tab:az_angles}
\begin{tabular}{ccc}
\hline
\hline
\textbf{True Az Angle (°)} & \textbf{Estimated Az Angle (°)} & \textbf{Error (°)} \\
\hline
$-20$ & $-13$ & $+7$ \\
$-17$ & $-23$ & $-6$ \\
$-14$ & $-7$  & $+7$ \\
$-11$ & $-16$ & $-5$ \\
$-8$  & $1$   & $+9$ \\
$-5$  & $-12$ & $-7$ \\
$-2$  & $3$   & $+5$ \\
$1$   & $-7$  & $-8$ \\
$4$   & $9$   & $+5$ \\
$7$   & $-1$  & $-8$ \\
$10$  & $17$  & $+7$ \\
$13$  & $5$   & $-8$ \\
$16$  & $21$  & $+5$ \\
$19$  & $28$  & $+9$ \\
\hline
\multicolumn{2}{c}{\textbf{RMSE}} & $7^\circ$ \\
\hline
\hline
\end{tabular}
\end{table}
The present prototype was calibrated and characterized for azimuth only; elevation scaling and calibration were not implemented, which currently limits elevation estimation. In addition, residual reflections and multipath in the laboratory environment introduce measurement uncertainty that contributes to the observed error. Both effects can be mitigated by performing the characterization in an anechoic chamber and by extending the calibration procedure to the elevation channel, which the architecture already supports through the dedicated $\Delta_{\text{el}}$ path.

\section{Conclusion}
\label{sec:conclusion}
This paper presented the design, fabrication, and experimental validation of a compact monopulse direction-finding receiver for the X-band. The receiver integrates a $2 \times 2$ antipodal Vivaldi array, which provides good impedance matching across \SIrange{8}{12}{\giga\hertz} with a simulated realised gain of approximately \SI{6.44}{dBi}, with a rat-race comparator network that synthesises the sum, azimuth-difference, and elevation-difference channels. The comparator maintains the matched response and the \SI{180}{\degree} phase balance at its difference ports required for unambiguous angle encoding, and the fabricated prototype preserves this behaviour across the band. An LTC5564 power-detector front-end, on-board digitisation, and a MATLAB graphical user interface together enable real-time monopulse-ratio computation and angle estimation. Azimuth direction finding was experimentally demonstrated at \SI{11}{\giga\hertz} over fourteen angles spanning \SIrange{-20}{19}{\degree}, yielding a root-mean-square error of \SI{7}{\degree} with the per-angle error bounded within $\pm\SI{9}{\degree}$. These results establish a passive, low-cost, and fully integrated receiver suitable for radar sensing, electronic warfare, and surveillance. Future work will extend the calibration to the elevation channel and characterise the system in an anechoic chamber to realise full two-dimensional operation.

\bibliographystyle{IEEEtran}

\end{document}